\documentstyle[12pt]{article}
\makeatletter
\newcommand{\bd}{\begin{document}}
\newcommand{\ed}{\end{document}}
\newcommand{\bc}{\begin{center}}
\newcommand{\bt}{\begin{tabbing}}
\newcommand{\ec}{\end{center}}
\newcommand{\et}{\end{tabbing}}
\newcommand{\be}{\begin{eqnarray}}
\newcommand{\ee}{\end{eqnarray}}
\newcommand{\nn}{\nonumber}
\newcommand{\eqn}{\global\def\theequation}
\def\kmu{K_L\to \mu\bar{\mu}}
\def\kplus{K^+\to \pi^+\nu\bar{\nu}}
\def\knn{K_L\to \pi^0 \nu\bar{\nu} }
\def\knu4{K_L\to \pi^+\pi^- \nu\bar{\nu}}
\def\bbar{B^0_d - \bar{B}_d^0}
\def\kee{K_L\to\pi^0 e^+ e^-}

\def\figcap{\section*{Figure Captions\markboth
     {FIGURECAPTIONS}{FIGURECAPTIONS}}\list
     {Figure \arabic{enumi}:\hfill}{\settowidth\labelwidth{Figure 999:}
     \leftmargin\labelwidth
     \advance\leftmargin\labelsep\usecounter{enumi}}}
\let\endfigcap\endlist \relax
\def\reflist{\section*{References\markboth
     {REFLIST}{REFLIST}}\list
     {[\arabic{enumi}]\hfill}{\settowidth\labelwidth{[999]}
     \leftmargin\labelwidth
     \advance\leftmargin\labelsep\usecounter{enumi}}}
\let\endreflist\endlist \relax

\renewcommand{\thefootnote}{\alph{footnote}}
\begin{document}
\tolerance=10000
\begin{titlepage}
\begin{flushright}
{\normalsize     NHCU-HEP-94-04 }
\end{flushright}
 \null
 \vskip 0.5in
\begin{center}
 \vspace{.15in}
{\Large {\bf  CP CONSERVING AND VIOLATING
\\
 \vspace{.10in}
 CONTRIBUTIONS TO $\knu4$
    }}
  \par
 \vskip 2.5em
 {\large
  \begin{tabular}[t]{c}
        {\bf C.~Q.~Geng$^{a}$, I.~J.~Hsu$^{b}$ and Y.~C.~Lin$^b$}
\\
\\
       $^a$ Department of Physics, National Tsing Hua University \\
    Hsinchu, Taiwan, Republic of China \\
and\\
       $^b$ Department of Physics, National Central University \\
           Chung-Li, Taiwan, Republic of China\\
   \end{tabular}}
 \par \vskip 5.0em
 {\large\bf Abstract}
\end{center}
\setlength{\baselineskip}{5ex}

We study both CP conserving and violating contributions
to the decay $\knu4$. We find that the decay branching ratio
is dominated by the CP conserving part.
In the standard CKM model, we estimate that for $m_t\sim 174\ GeV$,
the branching ratio due to the CP conserving (violating)
contributions can be as large as
$4.4 \times 10^{-13}$ ($1.0\times 10^{-14}$).

\vspace{1.0in}
\medskip
{\footnotesize}
\vfill

\end{titlepage}

\setlength{\baselineskip}{5ex}
\pagestyle{plain}
\pagenumbering{arabic}
\setcounter{page}{2}

\bc
{\bf I. INTRODUCTION}
\ec

With the prospect of a new generation of ongoing kaon experiments
a number of rare kaon decays have been suggested to test the
Cabibbo-Kobayashi-Maskawa (CKM) \cite{ckm} paradigm: Quarks of different
flavor are mixed in the charged weak currents by means of an
unitary matrix $V$. However it is sometimes a hard task to extract
the short-distance contribution, which depends on the CKM matrix, because
of large theoretical uncertainties in the long-distance contribution to
the decays \cite{review}.
To avoid this difficulty, much of recent theoretical as well as
experimental attention has been on searching for
the two modes: $\kplus$ and $\knn$. It is believed that these two decays are
free of long-distance and other theoretical uncertainties \cite{ld,wise}.

It has been shown that the decay branching ratio of $\kplus$
is at the level of $10^{-10}$  \cite{bg,others_k}
arising dominated from
the short-distance loop contributions containing virtual charm and top quarks.
This decay is a CP conserving process and probably the cleanest one,
in the sense of theoretical uncertainties,
to study the absolute value of the CKM element $V_{td}$.
The current experimental limit is
$Br(\kplus)_{expt}\leq 5\times 10^{-10}$ \cite{E787}
given by the ongoing E787 experiment at BNL.
It is expected that the experiment will reach the standard model
predicted level in a few years.
On the other hand,
the decay $\knn$ depending on the imaginary part of $V_{td}$ is
a CP violating process \cite{knn}
and offers a clear information about the origin
of CP violation. In the standard model, it is dominated by the Z-penguin
and W-box loop diagrams with virtual top quark.
But there has been no dedicated experimental search
for this decay yet.
Although there are several interesting proposals
to study this mode at the next round KEK and FNAL experiments \cite{kek_p},
the experimental sensitivities can only be around $10^{-9}$, whereas
the decay branching ratio in the CKM model is at the level of
$10^{-11}$ \cite{bg}.
 From an experimental point of view very challenging
efforts are necessary to perform the experiments.
This is because all the final state particles are neutral and
the only detectable particles are $2\gamma$'s from $\pi^0$.

In this paper, we examine the decay $\knu4$. Like the decays of
$\kplus$ and $\knn$, we expect that this mode is also a clean
one due to the absence of photon intermediate states.\footnote{We
note that the decay of $\knu4$ is different from
that of $K_L\to\pi^+\pi^-e^+e^-$ in which
it is dominated by the long distance due to the photon intermediate
states \cite{Sehgal}.}
Moreover, in contrast with $\knn$,
it contains two charge particles $\pi^+$ and $\pi^-$ in the final states
and could be relatively easy to do an experiment \cite{Bryman}.  Therefore,
it should be interesting to give a theoretical analysis
on this decay to see whether it could be tested experimentally
in future kaon facilities.

The paper is organized as follows. In Sect. II, we
study the decay rate of $\knu4$ from the short and long
distance contributions.
We present our numerical results in Sect. III. The conclusions
are given in Sect. IV.

\bc
{\bf II. DECAY RATES}
\ec

We start by writing the decay as
\be
K_L(p_K)\to \pi^+(p_+)\pi^-(p_-)\nu(k_+)\bar{\nu}(k_-)
\ee
where $p_K\,,\ p_+\,,\ p_-\,,\ k_+$ and $k_-$ are the four-momenta of
$K_L\,,\ \pi^+\,,\ \pi^-\,,\ \nu$ and $\bar{\nu}$, respectively.
Similar to the decays of $\kplus$ and $\knn$,
the short distance contributions, arising
 from the box and penguin loop diagrams with virtual charm
and top quarks, dominate the decay branching ratio of
$\knu4$. The effective interaction relevant for the process
is given by \cite{IL}
\be
{\cal L}_{eff} &= & {G_F\over \sqrt{2}}{\alpha\over 4\pi\sin^2\theta_W}
\sum_{i=c,t} V_{is}^*V_{id}\eta_i C_{\nu}(x_i)\: \bar{s}\gamma_{\mu}
(1-\gamma_5)d\;\bar{\nu}_l\gamma^{\mu}(1-\gamma_5)\nu_l
\ee
where $\eta_c\simeq 0.71$ and $\eta_t\simeq 1$ are the QCD correction
factors \cite{qcd},
$x_i=m_i^2/M_W^2$
and
\be
C_{\nu}(x_i) &= & {x_i\over 4}\left[{3(x_i-2)\over (x_i-1)^2}\ln x_i\;+\;
{x_i+2\over x_i-1}\right]\,.
\ee

To obtain the matrix element,
we follow the analysis of $K_{l4}$ decays by Pais and Treiman \cite{Pais}.
We define the following combinations of four-momenta:
\be
P\:=\: p_++p_-\,, && Q \:=\:p_+-p_-\,, \\
\nn
L\:=\: k_++k_-\,, && N \:=\:k_+-k_-\,.
\ee
Similar to $K_{l4}$ decays \cite{Pais},
the decay $\knu4$ can be kinematically parametrized
by five variables: $s_{\pi}=P^2$, the invariant mass of $\pi^+\pi^-$ pair;
$s_{\nu}=L^2$, the invariant mass of $\nu\bar{\nu}$ pair;
$\theta_{\pi}$, the angle between $\vec{p}_+$ and $\vec{L}$ as measured
in the $\pi^+\pi^-$ c.m. frame;
$\theta_{\nu}$, the angle between $\vec{k}_+$ and $\vec{P}$ as measured
in the $\nu\bar{\nu}$ c.m. frame; and $\phi$, the angle between the normals
to the $\pi^+\pi^-$ and $\nu\bar{\nu}$ planes.
The ranges of the variables are \cite{Ecker}:
\be
\nn
4M_{\pi}^2\leq & s_{\pi} &  \leq M_K^2\,, \\
\nn
0 \leq & s_{\nu} &  \leq (M_K-\sqrt{s_{\pi}})^2 \,,\\
\nn
0 \leq & \theta_{\pi}, & \theta_{\nu}\: \leq\: \pi \,,\\
0 \leq & \phi & \leq 2\pi\,,
\ee
respectively.
For the hadronic matrix element, we use the standard parametrization:
\be
<\pi^+\pi^-|\bar{s}\gamma_{\mu}(1-\gamma_5)d|K^0>
& = & {i\over M_K}\left[FP_{\mu}+GQ_{\mu}
+i{H\over M_K^2}\epsilon_{\mu\nu\rho\sigma}
L^{\nu}P^{\rho}Q^{\sigma}\right]
\ee
where the form factors $F\,,\ G$ and $H$ can be related by isospin
to the corresponding form factors in the matrix element of
$<\pi^+\pi^-|\bar{s}\gamma_{\mu}(1-\gamma_5)u|K^+>$
in $K_{l4}$ decay. These form factors have been evaluated in ChPT at order
$p^4$ \cite{Bijens,Gasser}. It is found that
\be\nn
F&=& G\:=\: {M_K\over f_{\pi}}\,,\\
H &=& {M_K^3\over 2\pi^2 f_{\pi}^3}
\ee
with $f_{\pi}=130\ MeV$.
 From Eqs. (2) and (6), we obtain the amplitude of the decay
$K^0\to\pi^+\pi^-\nu\bar{\nu}$ for each neutrino flavor as
\be\nn
A(K^0\to\pi^+\pi^-\nu\bar{\nu})&=&
- {G_F\over \sqrt{2}}{\alpha\over 4\pi\sin^2\theta_W}
\sum_{i=c,t} V_{is}^*V_{id}\eta_i C_{\nu}(x_i)\:
{i\over M_K}[F\;P_{\mu}+G\;Q_{\mu} \\
&&
+i{H\over M_K^2}\epsilon_{\mu\nu\rho\sigma}
L^{\nu}P^{\rho}Q^{\sigma}]
\;\bar{\nu}_l\gamma^{\mu}(1-\gamma_5)\nu_l \,.
\ee
With the CPT theorem and
$K_L\simeq K_2+\epsilon K_1\simeq(K^0-\bar{K}^0)/\sqrt{2}i$,
we find
\be\nn
A(K_L\to\pi^+\pi^-\nu\bar{\nu}) &=&
{G_F\over \sqrt{2}}{\alpha\over 4\pi\sin^2\theta_W}
{\sqrt{2} \lambda \over M_K}
\left\{ iG\;Q_{\mu}\left[-A^2 \lambda^4\eta C_{\nu}(x_t)\right]\right. \\
\nn
&&+\left(F\;P_{\mu}
+i{H\over M_K^2}\;\epsilon_{\mu\nu\rho\sigma}L^{\nu}P^{\rho}Q^{\sigma}\right)
\left[ \eta_cC_{\nu}(x_c) \right. \\
&&
\left. +A^2\lambda^4(1-\rho)C_{\nu}(x_t)]\right\}
\;\bar{\nu}_l\gamma^{\mu}(1-\gamma_5)\nu_l \,
\ee
where $\lambda=0.22$ is the Cabibbo angle,
$A\,,\ \rho$ and $\eta$ are the parameters in the Wolfenstein
parametrization \cite{wolf}
of the CKM matrix and we have ignored the contribution
from $K_1$ part because of the smallness of $\epsilon$ parameter.
In Eq. (9), the terms proportional to $F$ and $H$, which represent
$I=0$ $s$-wave and $I=1$ $p$-wave for the $\pi^+\pi^-$ system,
are CP conserving and that to $G$, $I=1$ $p$-wave,
is CP violating.

To write the partial decay rate for (1), it is convenient to introduce
the following combination of kinematic factors and form factors:
\be\nn
F_1 &=& -iF\;X \left[\eta_cC_{\nu}(x_c)
+A^2\lambda^4(1-\rho)C_{\nu}(x_t)\right]\\
\nn     && - \sigma_{\pi}(P\cdot L)\cos\theta_{\pi}
G\left[A^2 \lambda^4\eta C_{\nu}(x_t)\right]\,, \\
\nn
F_2 &=& -\sigma_{\pi}(s_{\pi}s_{\nu})^{1\over 2}
G\left[A^2 \lambda^4\eta C_{\nu}(x_t)\right]\,, \\
F_3 &=&
-\sigma_{\pi}X(s_{\pi}s_{\nu})^{1\over 2}
{iH\over M_K^2}\left[\eta_cC_{\nu}(x_c)
+A^2\lambda^4(1-\rho)C_{\nu}(x_t)\right]\,,
\ee
where
\be
\sigma_{\pi}\:=\:\left(1-{4M_{\pi}^2\over s_{\pi}}\right)^{1/2}\,,\
X\:=\: \left[(P\cdot L)^2-s_{\pi}s_{\nu}\right]^{1/2}\,,\
P\cdot L\:=\:{1\over 2}(M_K^2-s_{\pi}-s_{\nu})\,.
\ee
The differential decay rate is
\be
d^5\Gamma = {G_F^2\over 2^{12}\pi^6M_K^5}
\left({\alpha\sqrt{2}\lambda\over 4\pi\sin^2\theta_W}\right)^2
X\sigma_{\pi}I(s_{\pi},s_{\nu},\theta_{\pi},\theta_{\nu},\phi)
ds_{\pi}ds_{\nu}d\cos\theta_{\pi}d\cos\theta_{\nu}d\phi.
\ee
The dependence of $I$ on $\theta_{\nu}$ and $\phi$ is given by
\be\nn
I &=& I_1+I_2\cos 2\theta_{\nu}+I_3\sin^2\theta_{\nu}\cos 2\phi
+I_4\sin 2\theta_{\nu}\cos\phi+I_5\sin\theta_{\nu}\cos\phi \\
&& +I_6\cos\theta_{\nu}+I_7\sin\theta_{\nu}\sin\phi
+I_8\sin 2\theta_{\nu}\sin\phi
+I_9\sin^2\theta_{\nu}\sin 2\phi \;,
\ee
where $I_1,\cdots, I_9$ depend on $s_{\pi}\,,\ s_{\nu}$, and $\theta_{\pi}$.
By integrating over the angles, $\theta_{\nu}$ and $\phi$, we obtain
\be\nn
I(s_{\pi},s_{\nu},\theta_{\pi}) & =&
4\pi \left[ I_1-{1\over 3} I_2\right] \\
&=& {4\pi\over 3}\left[ |F_1|^2+(|F_2|^2+|F_3|^2)\sin^2\theta_{\pi}\right] \,,
\ee
where we have used the formulas for the form factors $I_1$ and $I_2$ given by
\be\nn
I_1 &= & {1\over 4}\left[|F_1|^2+{3\over 2}(|F_2|^2+
|F_3|^2)\sin^2\theta_{\pi}\right]\,, \\
I_2 &= & -{1\over 4}\left[|F_1|^2-{1\over 2}(|F_2|^2+
|F_3|^2)\sin^2\theta_{\pi}\right]\,.
\ee
Combining Eqs. (10)-(14), we get the differential decay rate of
$\knu4$ for three generations of neutrinos as follows
\be
{d^3\Gamma\over ds_{\pi}ds_{\nu}d\cos\theta_{\pi}}
&=& \left({d^3\Gamma\over ds_{\pi}ds_{\nu}d\cos\theta_{\pi}}\right)_{CPC}
\: + \: \left({d^3\Gamma\over ds_{\pi}ds_{\nu}d\cos\theta_{\pi}}\right)_{CPV}
\ee
with
\be\nn
\left({d^3\Gamma\over ds_{\pi}ds_{\nu}d\cos\theta_{\pi}}\right)_{CPC}
 &=& {G_F^2\over 2^{10}\pi^5M_K^5}
\left({\alpha\sqrt{2}\lambda\over 4\pi\sin^2\theta_W}\right)^2
X^3\sigma_{\pi}\left(F^2+\sigma^2_{\pi}s_{\pi}s_{\nu}
{H^2\over M^4_K}\sin^2\theta_{\pi}\right) \\
&& \cdot\left(\eta_cC_{\nu}(x_c) +A^2\lambda^4(1-\rho)C_{\nu}(x_t)\right)^2
\ee
and
\be\nn
\left({d^3\Gamma\over ds_{\pi}ds_{\nu}d\cos\theta_{\pi}}\right)_{CPV}
 &=& {G_F^2\over 2^{10}\pi^5M_K^5}
\left({\alpha\sqrt{2}\lambda\over 4\pi\sin^2\theta_W}\right)^2
X\sigma_{\pi}^3
\left(X^2\cos^2\theta_{\pi}+s_{\pi}s_{\nu}\right)
G^2\\
&& \cdot
\left(A^2 \lambda^4\eta C_{\nu}(x_t)\right)^2\,,
\ee
corresponding to the CP conserving and violating contributions, respectively.
As comparisons, we give the branching ratios for $\kplus$ and $\knn$ from the
short distance contributions:
\be\nn
Br(\kplus) &=& {3\over 2}\left({\alpha\over 2\pi\sin^2\theta_W}\right)^2
Br(K^+\to\pi^0e^+\nu) \\
\nn
&& \cdot\left[\left(\eta_cC_{\nu}(x_c) +A^2\lambda^4(1-\rho)C_{\nu}(x_t)
\right)^2+\left(A^2 \lambda^4\eta C_{\nu}(x_t)\right)^2\right] \,,\\
\nn
Br(\knn) &=& {3\over 2}\left({\alpha\over 2\pi\sin^2\theta_W}\right)^2
Br(K^+\to\pi^0e^+\nu){\tau (K_L)\over \tau (K^+)} \\
&& \cdot\left (A^2 \lambda^4\eta C_{\nu}(x_t)\right)^2\,,
\ee
where $Br(K^+\to\pi^0e^+\nu)=0.048$.
It is interesting to note that the $\knu4$ decay rate of the
CP violating part in Eq. (18)
has a similar CKM dependence as $\knn$ in Eq. (19) while that of the CP
conserving part in Eq. (17) is somewhat different from the CP conserving
decay of $\kplus$.

The long distance contribution can be calculated in the framework of chiral
perturbation theory. There are three kinds of terms which contribute
to the process of interest, $L^{\Delta S=1}_{(2)}$ \cite{wise}, reducible
anomaly ($L_{r.a.}$)
and direct anomaly ($L_{d.a.}$) \cite{bijn92}. $L^{\Delta S=1}_{(2)}$ is the
weak chiral lagrangian of $O(p^2)$
\be
L^{\Delta S=1}_{(2)}
& = & {G_8 f_{\pi}^4\over 4} Tr \lambda_6 D_{\mu}U^\dagger
D^{\mu}U
\ee
where
\be
U & = & \exp \left({i\sqrt{2}\over f_{\pi}} \phi^a
\lambda^a\right)
\ee
is the nonlinear realization of the octet meson fields and
\be
D_{\mu}U & = &  \partial_{\mu}U - i r_{\mu}U + iU l_{\mu}
\ee
is the covariant derivative with
\be
l_{\mu} & = & {g\over \cos \theta_w}Z_{\mu}\left( Q - {\xi\over 6}
              -\sin^2 \theta_w Q\right), \nn \\
r_{\mu} & = & {g\over \cos \theta_w}Z_{\mu}\left(
              -\sin^2 \theta_w Q\right).
\ee
The overall normalization $G_8$ is determined by the
amplitude of $K\rightarrow\pi\pi$ and the numerical value
is $9 \times 10^{-6}\ GeV^{-2}$. The matrix $Q$ is the quark charge matrix,
$Q = diag( 2/3,-1/3,-1/3)$, which characterizes the E.M. current coupling of
$Z$. The
parameter $\xi$ inside the left handed current $l_{\mu}$ is the coefficient for
the singlet current coupling of Z, and it is of unity in the limit of nonet
symmetry. Note that we have different identification for $l_{\mu}$ and
$r_{\mu}$ than those in \cite{wise}.
The reducible anomaly arises from the kind of diagrams starting with a $K-\pi$
(or $K-\eta$) weak transition induced by $L^{\Delta S=1}_{(2)}$, then
followed by
a $\pi$ (or $\eta$) pole and
ended by an anomaly vertex derived from $L_{W.Z.W}$.
The relevant pieces to our calculation in $L_{W.Z.W}$ are given by
\be
L_{W.Z.W}& = &
- {i\over 16 \pi^2}Tr
\epsilon_{\mu\nu\alpha\beta}L^{\mu}L^{\nu}L^{\alpha}l^{\beta}
+ {i\over 16 \pi^2}Tr
\epsilon_{\mu\nu\alpha\beta}R^{\mu}R^{\nu}R^{\alpha}r^{\beta}
\ee
where
\be
L_{\mu} & = & i U^{\dagger}\partial_{\mu}U ,\nn\\
R_{\mu} & = & i U\partial_{\mu}U^{\dagger}.
\ee
The direct anomaly can be understood as the bosonization of
the product of the left handed currents arising from $L^{\Delta
S=1}_{(2)}$ and $L_{W.Z.W.}$. It reads
\be
L_{d.a.} & = & {G_8 f_{\pi}^2 \over 32 \pi^2}\left\{ 2a_1 i
\epsilon^{\mu\nu\alpha\beta}Tr\lambda_6 L_{\mu} Tr
L_{\nu}L_{\alpha}L_{\beta}\right. \nn\\
& & + a_2 Tr \lambda_6 [U^{\dagger}F^{\mu\nu}_R U,
L_{\mu}L_{\nu}] + 3a_3Tr\lambda_6 L_{\mu}Tr (F^{\mu\nu}_L +
U^{\dagger}F^{\mu\nu}_R U)L_{\nu} \nn\\
& & \left. + a_4 Tr\lambda_6 L_{\nu}Tr (F^{\mu\nu}_L -
U^{\dagger}F^{\mu\nu}_R U)L_{\nu}\right\}
\ee
where $F^{\mu\nu}_{R,L}$ are the field strengths associated
with the fields $r_{\mu}$ and $l_{\mu}$ correspondingly.
The coefficients $a_i$ are in principle of order one and
they can be extracted from the anomalous radiative decay modes of koan.
In terms of the kinetics variables defined before, the decay amplitude
resulting from long distance effect is given by
\be
A_L(K_L\rightarrow\pi^+\pi^-\nu\bar{\nu}) & = &
-{i \sqrt{2} g^2 G_8\over 32 \pi^2 f_{\pi}M_Z^2\cos^2\theta_w}
\bar{\nu}_l\gamma_{\mu}
(1-\gamma_5)\nu_l
\{\epsilon^{\mu\nu\alpha\beta}L_{\nu}P_{\alpha}Q_{\beta}
\nn\\
&&
\left[2(3a_1-3a_3-a_4)+2\sin^2\theta_w(a_2+2a_4) +
\xi{m^2_K \over m^2_K - m^2_{\pi}}
\right]\nn\\
& & -8i\pi^2 f^2_{\pi}(1-2\sin^2\theta_w)(P^{\mu}+L^{\mu})
\},
\ee
and the corresponding differential decay rate is then given by
\be
\left({d^3\Gamma\over ds_{\pi}ds_{\nu}d\cos\theta_{\pi}}\right)_L & = &
{g^4 G_8^2 \sigma_{\pi}\over 2^{19}\pi^9 M_Z^4
M_K^3}\cos^4\theta_{\pi}
\left\{ 16\pi^2 f_{\pi}^4 (1-2\sin^2\theta_w)^2\right.
\nn\\
&&
+ \sigma_{\pi}^2 \sin^2
\theta_w s_{\pi}s_{\nu}
\left[2(3a_1-3a_3-a_4) \right.
\nn\\
&&
\left.\left. +2\sin^2\theta_w(a_2+2a_4) +
\xi m^2_K/(m^2_K - m^2_{\pi}) \right]^2\right\}.
\ee
\bc
{\bf III. NUMERICAL RESULTS}
\ec

The validity of relating $m_t$ to the decay rate depends upon the negligibility
of long distance contribution. Therefore it is important to learn the
branching ratio arising from the long distance effect.
Due to the absence of $m_t$ in the amplitude, the decay rate of long distance
contribution
is relatively suppressed by at least two orders. Numerically we find
\be
Br( L_{d.a.} + L_{r.a.}) & \sim & 4.7 \times 10^{-20},\nn\\
Br(L^{\Delta S=1}_{(2)}) & = & 5.0 \times 10^{-18}.
\ee
As we shall see below it is safe to ignore the long distance effect and
we shall concentrate on the short distance effect only in the following
analysis of decay rate.

To estimate the CP conserving and
violating decay rates in (17) and (18), we need to find out the allowed values
for the CKM parameters $A\,,\ \rho$ and $\eta$, constrained by the
experimental measurements such as
$\epsilon$, the CP violation parameter in $K\to \pi\pi$;
$x_d$, the $B_d^0-\bar{B}_d^0$ mixing; and the ratios
$|V_{cb}/V_{us}^2|$ and $|V_{ub}/V_{cb}|$ of the CKM elements.
We use the same fitting procedure and the necessary equations
in Refs. \cite{bg,gt}.
In the fits, we take the updated values $|V_{cb}|=0.041\pm0.005$
and  $|V_{ub}/V_{cb}|=0.080\pm0.025$ and $f_B=200\pm50\ MeV$.

Integrating over all the variables in Eqs. (17) and (18),
we can examine the decay rates
for both CP conserving and violating
parts which depend on the top quark mass and the CKM parameters.
In Figures 1a and 1b, we plot the
branching ratios of CP conserving and violating
contributions to $\knu4$ as a function of the top quark mass,
showing the lower and higher values allowed at $90\%$ C.L.,
where $Br(\knu4)=\Gamma(\knu4)/\Gamma(K_L\to all)$.
We also show the corresponding decay branching ratios of $\kplus$ and
$\knn$ in Figures 1c and 1d by using Eq. (19), respectively.
 From the figure,
we see that the CP conserving part of the branching ratio
is much larger than that of CP violating one.
Clearly, measuring the decay rate will not give us information
on the CP violation.
For $150\leq m_t\leq 200\ GeV$, we find
\be\nn
1.1\times 10^{-13}\leq & Br(\knu4)_{CPC} & \leq 5.0\times 10^{-13}\,,\\
5.0\times 10^{-16}\leq & Br(\knu4)_{CPV} & \leq 1.1\times 10^{-14}\,.
\ee

We now study the differential decay spectrum in terms of $s_{\pi}$
($\theta_{\pi}$) by integrating over $s_{\nu}$ and $\theta_{\pi}$
($s_{\pi}$  and $s_{\nu}$) in Eqs. (17) and (18) to see whether
we would distinguish the CP conserving and violating parts.
We define
the normalized invariant mass of $\pi^+\pi^-$ as
$x=s_{\pi}/M_K^2$.
To illustrate the shapes
of the spectra between the CP conserving and violating cases,
we choose $m_t\sim 160\ GeV$ and the CKM parameters
$A\sim 1\,,\ \rho\sim -0.2$ and $\eta\sim 0.4$.
We plot the differential branching ratios
$dBr(\knu4)/dx^{1\over 2}$ vs
$x^{1\over 2}$ and $dBr(\knu4)/d\cos\theta_{\pi}$ vs
$\cos\theta_{\pi}$ in Figures 2 and 3,
respectively.
As shown in Figure 2, the CP conserving and violating
spectra of
$dBr(\knu4)/dx^{1\over 2}$
have similar shapes and are dominated by small values
of $s_{\pi}$. However, in Figure 3, as expected,
$\left[dBr(\knu4)/d\cos\theta_{\pi}\right]_{CPV}$ becomes maximum
when $\theta_{\pi}$ is close to $0$ or $\pi$ and to minimal
when it reaches $\pi/2$ whereas
$\left[dBr(\knu4)/d\cos\theta_{\pi}\right]_{CPC}$ does the opposite way.
Unfortunately, the values of
the CP violating one around $\theta_{\pi}=0,\pi$
may be still too small to be tested.

\bc
{\bf IV. CONCLUSIONS}
\ec

We have studied both short and long distance contributions to the decay of
$\knu4$. We have demonstrated that the long distance effect to the decay rate
is negligible small.
We have shown that the branching ratio of the decay is dominated
by the CP conserving part. With the updated CKM parameters, we find
that the decay branching ratio is predicted to be $(1- 5) \times
10^{-13}$ for $m_t\leq 200\ GeV$, which could be accessible to
experiments at future kaon facilities.
The CP violating contribution to the branching ratio seems
impossible to be measured in experiments.
However, it is, in principle, to distinguish the CP conserving
and violating contributions by measuring
the spectra of the $\theta_{\pi}$ angular dependence
of the differential decay rates.

\begin{center}
{\bf ACKNOWLEDGMENTS}
\end{center}

We thank Professor D. Bryman for
suggesting us to study this decay mode.
We thank  D. Bryman, H. Y. Cheng, G. Ecker, T. Inagaki,
Y. Kuno, W. J. Marciano and B. Winstein for useful discussions.

This work was supported in part
by the National Science Council of
Republic of China
under Grant No. NSC-83-0208-M-007-118 (C.Q.G) and
NSC-83-0208-M-008-009 (I.-J.H and Y.C.L).

\newpage

\begin{figcap}

\item
Allowed branching ratios for
(a) CP conserving contribution to $\knu4$;
(b) CP violating contribution to $\knu4$;
(c) $\kplus$; and (d) $\knn$ as functions of $m_t$
at $90\%$ C.L.
\item The differential decay spectrum of $d\Gamma(\knu4)/dx^{1\over 2}$
as a function of $x^{1\over 2}=\sqrt{s_{\pi}}/M_K$ with
$m_t=160\ GeV\,,\ A\sim 1.0\,,\ \rho\sim -0.2$ and
$\eta\sim 0.4$.
\item The differential decay spectrum of $d\Gamma(\knu4)/d\cos\theta_{\pi}$
as a function of $\cos\theta_{\pi}$.
Legend is the same as in Figure 2.

\end{figcap}

\ed

